\documentclass[mypaper,8pt,twoside]{CoAst}
\usepackage{epsf,graphicx,fancyhdr,sfmath}
\renewcommand{\chaptermark}[1]{\markboth{#1}{}}

\newcommand{\kopf}{\small\itshape Comm.\ in Asteroseismology, N$^{\textsf{\underline{o}}}$ 159, 2009\\
Proceedings of the JENAM 2008 Symposium N$^{\textsf{\underline{o}}}$~4:
Asteroseismology and Stellar Evolution}

\newcommand{\Authors}[1]{\begin{center}\normalsize\bf\sf #1 \end{center}}

\renewcommand{\author}[1]{\begin{center}\normalsize\bf\sf #1 \end{center}}
\newcommand{\Address}[1]{\begin{center}\small\sf #1 \end{center}}

\newcommand{\References}[1]{\begin{flushleft}{\large References\\}\vspace*{2mm}\small #1 \end{flushleft}}

\newcommand{\chapterCoAst}[2]{\chapter[\sf\normalsize #1\\ \footnotesize \hspace*{5mm}by #2 \sf\normalsize][]{#1\\}\rhead[\fancyplain{}{\sf\footnotesize \center{#1}}]{\fancyplain{}{\sffamily\thepage}}\lhead[\fancyplain{\kopf}{\sffamily\thepage}]{\fancyplain{\kopf}{\sf\footnotesize \center{#2}}}}

\renewcommand{\chaptername}{}

\newcommand{\acknowledgments}[1]{\vspace*{5mm}\noindent  \textbf{Acknowledgments.} #1}

\def\rfr{\smallskip\par\noindent
        \hangindent=7truemm
        \hangafter=1}

\begin{document}
\sf

\chapterCoAst{Further progress on solar age calibration}%paper title and page heading for even pages
{G.\,Houdek, D.O.\,Gough} %page heading for odd pages
\vspace{-10pt}
\Authors{G.\,Houdek$^{1}$ and D.O.\,Gough$^{1,2}$} 
\vspace{-10pt}
\Address{
$^1$ Institute of Astronomy, Madingley Road, Cambridge CB3\,0HA, UK\\
$^2$ Department of Applied Mathematics and Theoretical Physics, Cambridge CB3\,0WA, UK
%$^2$ Department of Applied Mathematics and Theoretical Physics, Wilberforce Road, Cambridge CB3\,0WA, UK
}

%\vspace{-3pt}
%\noindent
%\begin{abstract}
%\vspace{-4pt}
%We recalibrate a standard solar model seismologically to estimate 
%the main-sequence age of the Sun. Our procedure differs from what 
%we have done in the past by removing from the observed frequencies
%a crude representation of the effect of hydrogen ionization and 
%the superadiabatic convective boundary layer. Our preliminary result
%is $t_\odot=5.63\pm0.02\,Gy$.
%\end{abstract}
%\Objects{Sun.}
\vspace{-8pt}
\section*{Introduction}
\vspace{-6pt}
Seismological calibration of solar models to estimate the age of the 
Sun necessarily depends predominantly on the frequencies of the 
lowest-degree modes which penetrate into the energy-generating core 
where the greatest evolutionary change in the stratification occurs. 
Most commonly this is accomplished by fitting the asymptotic formula
\vspace{-2pt}
\begin{equation}
\nu_{n,l}\sim\left[n+\frac{1}{2}l+\hat\epsilon
-\sum_{k=1}^K\left(\sum_{j=0}^kA_{k,j}L^{2j}\right)
\left(\frac{\nu_0}{\nu_{n,l}}\right)^{2k-1}\right]\nu_0\,,
\vspace{-2pt}
\end{equation}
to observed high-order frequencies $\nu_{{\rm o}n,l}$ of order $n$ and degree $l$
to determine the coefficients $\nu_0$, $\hat\epsilon$, and $A_{k,j}$; here
$L=l+\frac{1}{2}$. The most $l-$sensitive terms, at each degree $2k\!-\!1$,
namely $A_{k,k}$, are, on the whole, the most sensitive to core
conditions, and the least sensitive to the structure of the envelope
(cf. Houdek \& Gough 2007b). Therefore it is one or more of these 
that are the best determinants of stellar age.
Eq.\,(1) is valid only if $l\ll n$ and $n$ is large, such that the spatial scale 
of variation of the equilibrium state is everywhere much greater than the 
inverse vertical wavenumber of the mode. But that condition is not actually
satisfied in the Sun: there is small-scale variation associated with ionization
of abundant elements and the near discontinuity in low derivatives of the
density at the base of the convection zone, which we call acoustic glitches, and
which add the components $\nu_{{\rm g}n,l}$ to $\nu_{n,l}$ that are in general 
oscillatory with respect to $n$. Ignoring these components introduces systematic 
errors into a straightforward fitting of Eq.\,(1) to $\nu_{{\rm o}n,l}$, errors 
that are evident in the undulatory age
estimates as the limits of the frequency range adopted for the fitting are varied
(Gough 2001). In an attempt to obviate these errors, Houdek \& Gough (2007a, 2008)
estimated the glitch components $\nu_{{\rm g}n,l}$ by fitting to second differences 
(with respect to $n$) of the observed frequencies an asymptotic formula designed to 
represent the base of the convection zone and the two ionization zones of helium. In
reality there is also an upper-glitch component, produced by the ionization of hydrogen 
and the upper superadiabatic boundary layer of the convection zone, which appears 
to be difficult to model in a reliable manner; when fitting the second differences
Houdek \& Gough (2007a) represented that component, coupled with the second differences
of Eq.~(1), somewhat arbitrarily as a series $P(\nu_{n,l})$ of inverse powers of
$\nu_{n,l}$. Because the upper glitch component is relatively smooth, they subsequently
tacitly regarded it as being included in the smooth asymptotic expression (1) by
adjusting the observed frequencies by only the component $\nu_{{\rm g}n,l}$.  Because the
upper glitch is quite close to the surface (partly in the evanescent zones of 
the modes), its influence on the eigenfrequencies is essentially independent of $l$, and
so should \hbox{not have materially affected the fitted coefficients $A_{k,j}$ with $j>0$.}
\vspace{-13pt}
\section*{Modification to the calibration procedure}
\vspace{-6pt}
In the work we report here we have tested the stability of the procedure by 
including in $\nu_{{\rm g}n,l}$ a representation of the upper-glitch component. 
To this end we summed the second-difference representation $P$ to obtain an estimate
of its contribution to the frequencies. There is some ambiguity in how one separates smooth
and glitch components near the surface, which is exhibited by the two undetermined 
constants of summation of the second differences; here we chose those constants 
by minimizing the error-weighted sum of the squares of the upper-glitch frequencies.
The outcome is plotted in Fig.~1 using BiSON data (e.g. Basu et\,al. 2007) 
up to degree $l=3$.

After fitting Eq. (1), with $K=3$, to the resulting glitch-adjusted observed 
frequencies, the coefficients $\nu_0$, $\hat\epsilon$ and $A_{k,0}$ were found 
to be naturally somewhat different from the results obtained without the 
upper-glitch adjustment. But the coefficients $A_{k,k}$ are similar. There 
is, however, a slight difference, which is evidently a product of an inadequacy of
the asymptotic formulae to reproduce precisely the observed frequencies of the Sun.
\vspace{-7pt}
\section*{Result}
\vspace{-7pt}
The result of the present model calibration against BiSON data (e.g. Basu et\,al. 2007) is
\vspace{-3pt}
\begin{equation}
t_\odot=4.63\pm0.02\,{\rm Gy}\,,
\vspace{-3pt}
\end{equation}
a value in fair accord with our previous estimates (Houdek \& Gough 2007b, 2008). The errors
quoted here come solely from the stated observed frequency errors, which we have assumed
to be statistically independent, and take no account of (systematic) errors in our procedure;
that the value (2) differs from our previous estimates by as much as 2.5$\,\sigma$ suggests that
such systematic errors could be present at a level at least as great as the random errors.
Our current value for the solar age is lower than the previous estimates by essentially this method,
although, in contrast to many earlier estimates, it remains greater than the age of Model\,S of
Christensen-Dalsgaard et al. (1996), which we used as our reference. It is also greater than that of
many, if not all, meteorites. We have not yet completed our investigation of the robustness of the
result, so we offer it still as a preliminary estimate.
%==================================================
\begin{figure}
\centering
\mbox{
\begin{minipage}[ht!]{6.3cm}
\includegraphics[width=\textwidth]{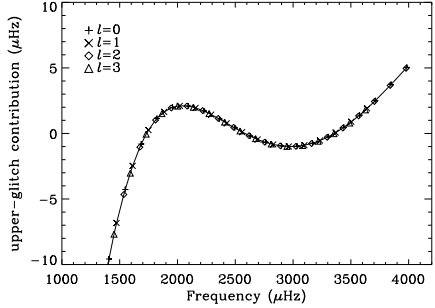}
\end{minipage}
\hspace{0.1cm}
\begin{minipage}[ht!]{4.62cm}
\vspace{-3.1mm}
\caption{Contribution of the upper-glitch component to the glitch frequencies 
$\nu_{{\rm g}n,l}$, $\!$obtained from the $\!$BiSON $\!$observations 
$\nu_{{\rm o}n,l}$ (Basu et\,al. 2007), as a function of $\nu_{{\rm o}n,l}$. 
It was obtained from summation of the series $P(\nu_{n,l})$, 
whose coefficients were determined from fitting to the second differences 
of the observed frequencies the second differences of 
an asymptotic formula representing the glitch components 
$\nu_{{\rm g}n,l}$ (Houdek \& Gough 2008). The upper-glitch component 
is produced by the ionization of hydrogen and the superadiabaticity of the
surface boundary layer.
}
\end{minipage}
}%\mbox
\label{fig:1}
\end{figure}
%==================================================

\vspace{-7pt}
\acknowledgments{
GH is grateful for support by the Science and Technology Facilities Council.
}
%\vspace{-1pt}
\References{
\vspace{-6pt}
\rfr Basu S., Chaplin, W.~J., Elsworth, Y., New, A.~M., Serenelli, G., Verner, G.~A. 2007, ApJ, 655, 660
\rfr Christensen-Dalsgaard, J., D{\"a}ppen. W., Ajukov, S.V., et~al.\ 1996, Science, 272, 1286
\rfr Gough, D.~O. 2001, in Astrophysical Ages and Timescales, von Hippel, T., Simpson, C., 
     Manset, N., eds, ASP Conf. Ser.\,245, Astron. Soc. Pac., San Francisco, p.\,31 
\rfr Houdek, G., Gough, D.~O. 2007a, MNRAS, 375, 861
\rfr Houdek, G., Gough, D.~O. 2007b, in: Unsolved Problems in Stellar Physics, 
     Stancliffe, R.~J., Dewi, J., Houdek, G., Martin, R.~G., Tout, C.~A., eds, 
     AIP Conf. Proc., AIP, New York, p.~219
\rfr Houdek, G., Gough, D.~O. 2008, in: The Art of Modelling Stars in the 21st Century, 
     Deng, L., Chan, K.~L., Chiosi C., eds, IAU Symp., Vol.\,252, CUP, Cambridge , p.~149
}

\end{document}